\documentclass[aps,prd,twocolumn,superscriptaddress]{revtex4}
\usepackage{epsfig,epsf}
\usepackage{amsmath}
\usepackage{amsthm}
\usepackage{amsfonts}
\usepackage{amssymb}
\usepackage{dsfont}
\usepackage{multirow}
\usepackage{appendix}
\usepackage{slashed}
\usepackage[active]{srcltx}
\usepackage{psfrag}

\begin{document}

\title{Axial-vector molecular structures $B_{c}^{\ast \pm} B_{c}^{\mp}$}
\date{\today}
\author{S.~S.~Agaev}
\affiliation{Institute for Physical Problems, Baku State University, Az--1148 Baku,
Azerbaijan}
\author{K.~Azizi}
\affiliation{Department of Physics, University of Tehran, North Karegar Avenue, Tehran
14395-547, Iran}
\affiliation{Department of Physics, Dogus University, Dudullu-\"{U}mraniye, 34775
Istanbul, T\"{u}rkiye}
\author{H.~Sundu}
\affiliation{Department of Physics Engineering, Istanbul Medeniyet University, 34700
Istanbul, T\"{u}rkiye}

\begin{abstract}
The axial-vector molecular structure $\mathcal{M}_{\mathrm{AV}}=(B_{c}^{\ast
+}B_{c}^{-} + B_{c}^{\ast - } B_{c}^{+})/2$ is explored in QCD sum rule
framework. The mass and current coupling of this compound are found by means
of the two-point sum rule method. It turns out that the mass of $\mathcal{M}%
_{\mathrm{AV}}$ is equal to $m=(12770 \pm 60)~\mathrm{MeV}$ making possible
its dissociations to pairs of $J/\psi \eta_b$, $\Upsilon \eta_c$, $%
B_{c}^{\ast +} B_{c}^{-}$, and $B_{c}^{\ast - } B_{c}^{+}$ mesons. The
partial widths of these decay modes contain the strong couplings at $%
\mathcal{M}_{\mathrm{AV}}$-meson-meson vertices which are calculated by
utilizing the three-point sum rule method. The full decay width of the
molecule $\mathcal{M}_{\mathrm{AV}}$ is formed due to these dominant
processes and amounts to $\Gamma[\mathcal{M}_{\mathrm{AV}}]=(93 \pm 14)~%
\mathrm{MeV}$. Our results for parameters $m$ and $\Gamma[\mathcal{M}_{%
\mathrm{AV}}]$ of the hadronic molecule $\mathcal{M}_{\mathrm{AV}}$ are
useful for experimental studies of various fully heavy four-quark mesons.
\end{abstract}

\maketitle

%%%%%%%%%%%%%%%%%%%%%%%%%%%%%%%%%%%%%%%%%%%%%%%%%%%%%%%%%%%%%%%%%

\section{Introduction}

\label{sec:Intro}
%%%%%%%%%%%%%%%%%%%%%%%%%%%%%%%%%%%%%%%%%%%%%%%%%%%%%%%%%%%

Four-quark exotic mesons, i.e., tetraquarks may exist in two configurations.
One of them is a diquark-antidiquark structure which is a bound or resonant
state of two colored diquark and antidiquark components. The second one is
compound built of two colorless conventional mesons. Both of these forms of
the tetraquarks are interesting objects for theoretical and experimental
studies. In fact, existence of the multiquark and hybrid hadrons, including
tetraquarks, are allowed by the quark-parton model and fundamental
principles of the quantum field theory of the strong interactions -- quantum
chromodynamics (QCD). Therefore, there is a necessity to explore such
structures to reveal their spectroscopic parameters and decay mechanisms,
find processes in which they may be observed. These analyses provide new
methods and schemes and/or lead to adaptation of well known approaches for
studying exotic hadrons. Theoretical predictions are also important for
planning and performing relevant experimental investigations.

The present work is devoted to a hadronic molecule made of two $B_{c}^{(\ast
)}$ mesons, therefore we restrict ourselves by considering relevant
theoretical and experimental results. It is worth to emphasize that heavy
hadronic molecules were objects of interesting theoretical investigations
\cite{Ding:2008mp,Albuquerque:2012rq,Chen:2015ata,
Karliner:2015ina,Liu:2016kqx,Chen:2017vai,PavonValderrama:2019ixb,Xin:2021wcr}%
. They still attract interest of researches, which explore different
characteristics of such systems \cite%
{Wang:2023bek,Braaten:2023vgs,Liu:2023gla,Liu:2024pio,Braaten:2024tbm,Yalikun:2025ssz}%
.

Over a few decades different experimental collaborations collected valuable
information on properties of hadronic resonances which may interpreted as
diquark-antidiquarks or hadronic molecules. The fully heavy $X$ resonances
were observed recently by LHCb, ATLAS, and CMS collaborations in di-$J/\psi $
and $J/\psi \psi ^{\prime }$ invariant mass distributions \cite%
{LHCb:2020bwg,Bouhova-Thacker:2022vnt,CMS:2023owd}. These resonances with
masses in the range $6.2-7.3~\mathrm{GeV}$ are presumably scalar $cc%
\overline{c}\overline{c}$ tetraquarks. Properties and possible internal
organizations of these resonances were studied in numerous papers \cite%
{Zhang:2020xtb,Albuquerque:2020hio,Wang:2022xja,Dong:2022sef,Dong:2020nwy,
Dong:2021lkh,Liang:2021fzr,Niu:2022vqp,Yu:2022lak,Kuang:2023vac} including
our publications \cite%
{Agaev:2023wua,Agaev:2023ruu,Agaev:2023gaq,Agaev:2023rpj} as well. Note that
the lightest resonance $X(6200)$ was interpreted there as an $\eta _{c}\eta
_{c}$ molecule, whereas structures $X(6900)$ and $X(7300)$ may be modeled as
admixtures of the different diquark-antidiquarks and $\chi _{c0}\chi _{c0}$,
$\chi _{c1}\chi _{c1}$ molecules, respectively.

The experimental discoveries proved that fully heavy tetraquarks are
accessible in ongoing experiments, thereby supported efforts devoted to
detailed analysis of tetraquarks containing only heavy $c$ and/or $b$
quarks. Hidden charm and bottom exotic mesons $bc\overline{b}\overline{c}$
are among such interesting particles. Various aspects of these tetraquarks
were intensively addressed in the literature (see Refs.\ \cite%
{Wu:2016vtq,Liu:2019zuc,Chen:2019vrj,Bedolla:2019zwg,Cordillo:2020sgc,
Weng:2020jao,Yang:2021zrc,Faustov:2022mvs,Hoffer:2024alv}, and references
therein). In our articles \cite{Agaev:2024wvp,Agaev:2024mng,Agaev:2024qbh},
we explored the scalar, axial-vector, and tensor diquark-antidiquark states $%
bc\overline{b}\overline{c}$ in the context of the QCD sum rule (SR) method \cite%
{Shifman:1978bx,Shifman:1978by} and computed their masses and full decay
widths.

The scalar hadronic molecule $\mathcal{M=}B_{c}^{+}B_{c}^{-}$ with the same
content $bc\overline{b}\overline{c}$ was investigated in Ref. \cite%
{Agaev:2025wdj}. It turned out that $\mathcal{M}$ is heavier than two-meson $%
\eta _{b}\eta _{c}$, $J/\psi \Upsilon $, and $B_{c}^{+}B_{c}^{-}$
thresholds, as a result, decays into these mesons are its kinematically
permitted decay channels. There are other mechanisms for decays of $\mathcal{%
M}$ into ordinary mesons \cite{Becchi:2020mjz, Becchi:2020uvq,Agaev:2023ara}%
. Thus, due to annihilations of $b\overline{b}$ and $c\overline{c}$ quarks
into $\overline{q}q$ and $\overline{s}s$ pairs, $\mathcal{M}$ readily decay
into $D_{(s)}^{(\ast )+}D_{(s)}^{(\ast )-}$ and $D^{(\ast )0}\overline{D}%
^{(\ast )0}$, as well as into $B^{(\ast )+}B^{(\ast )-}$ and $B_{(s)}^{(\ast
)0}\overline{B}_{(s)}^{(\ast )0}$ mesons.

The hadronic molecules $B_{c}^{(\ast )+}B_{c}^{(\ast )-}$ were also studied
in Ref.\ \cite{Liu:2023gla} using the coupled-channel unitary approach.
Results in this method depend on the parameter $\Lambda $ used to regularize
related integrals. In the case of the scalar structure $B_{c}^{+}B_{c}^{-}$
at $\Lambda =600~\mathrm{MeV}$ authors found for the pole position $E_{%
\mathrm{p}}=(12503.3-i126.8)~\mathrm{MeV}$. Stated differently, the mass of $%
B_{c}^{+}B_{c}^{-}$ is $12503.3~\mathrm{MeV}$ which is smaller than one
obtained in Ref. \cite{Agaev:2025wdj}. Nevertheless, in this picture\ the
molecule $B_{c}^{+}B_{c}^{-}$ is also unstable and can strongly decay into a
pair of $\eta _{b}\eta _{c}$ mesons.

In the current work, we explore the axial-vector molecule $\mathcal{M}_{%
\mathrm{AV}}=$ $(B_{c}^{\ast +}B_{c}^{-}+B_{c}^{+}B_{c}^{\ast -})/2$ with
spin-parities $J^{\mathrm{PC}}=1^{++}$ and compute the mass and full decay
width of this state. The mass $m$ and current coupling $\Lambda $ of $%
\mathcal{M}_{\mathrm{AV}}$ are calculated in the two-point sum rule
framework. As we shall see below, the mass of $\mathcal{M}_{\mathrm{AV}}$ is
large enough to ensure its dominant decays to $J/\psi \eta _{b}$, $\Upsilon
\eta _{c}$, $B_{c}^{\ast +}B_{c}^{-}$ and $B_{c}^{+}B_{c}^{\ast -}$ mesons.
The partial widths all of these decays are calculated by employing methods
of the three-point sum rule approach. This method is required to determine
the SR for the strong form factor at the relevant $\mathcal{M}_{\mathrm{AV}}$%
- meson-meson vertices. For instance, in the case of the decay $\mathcal{M}_{%
\mathrm{AV}}\rightarrow J/\psi \eta _{b}$ we find the sum rule for the form
factor $g_{1}(q^{2})$ responsible for strong interaction at the vertex $%
\mathcal{M}_{\mathrm{AV}}J/\psi \eta _{b}$ and evaluate its numerical value.
Computed by this way, the form factor $g_{1}(q^{2})$ at the mass shell of
the $J/\psi $ meson, i.e., at $q^{2}=m_{J/\psi }^{2}$ fixes the strong
coupling $g_{1}$ which is used to calculate the partial width of the process
under analysis.

This paper is divided into four sections: The mass and current coupling of
the axial-vector molecule $\mathcal{M}_{\mathrm{AV}}$ are found in Sec.\ \ref%
{sec:Mass}. The decays $\mathcal{M}_{\mathrm{AV}}\rightarrow J/\psi \eta
_{b} $, $\Upsilon \eta _{c}$, $B_{c}^{\ast +}B_{c}^{-}$ and $%
B_{c}^{+}B_{c}^{\ast -}$ are considered in Sec.\ \ref{sec:Widths1}. The full
decay width of the molecule $\mathcal{M}_{\mathrm{AV}}$ is determined in
this section as well. We make our concluding notes in the last Section\ \ref%
{sec:Conc}.

%%%%%%%%%%%%%%%%%%%%%%%%%%%%%%%%%%%%%%%%%%%%%%%%%%%%%%%%%%%%%%%%%

\section{Mass and current coupling of the molecule $\mathcal{M}_{\mathrm{AV}%
} $}

\label{sec:Mass}
%%%%%%%%%%%%%%%%%%%%%%%%%%%%%%%%%%%%%%%%%%%%%%%%%%%%%%%%%%%
The sum rules for the mass $m$ and current coupling $\Lambda $ of the
axial-vector hadronic molecule $\mathcal{M}_{\mathrm{AV}}$ can be extracted
from analysis of the correlation function%
\begin{equation}
\Pi _{\mu \nu }(p)=i\int d^{4}xe^{ipx}\langle 0|\mathcal{T}\{J_{\mu
}(x)J_{\nu }^{\dag }(0)\}|0\rangle .  \label{eq:CF1}
\end{equation}%
Here, $J_{\mu }(x)$ is the interpolating current for the molecule $\mathcal{M%
}_{\mathrm{AV}}$, whereas $\mathcal{T}$ denotes the time-ordered product of
currents.

In the hadronic molecule picture the interpolating current for $\mathcal{M}_{%
\mathrm{AV}}$ is
\begin{eqnarray}
J_{\mu }(x) &=&\frac{1}{2}\left\{ [\overline{b}_{a}(x)i\gamma _{5}c_{a}(x)][%
\overline{c}_{b}(x)\gamma _{\mu }b_{b}(x)]\right.  \label{eq:CR1} \\
&&\left. +[\overline{b}_{a}(x)\gamma _{\mu }c_{a}(x)][\overline{c}%
_{b}(x)i\gamma _{5}b_{b}(x)]\right\} ,
\end{eqnarray}%
where $a$, $b$ are the color indices.

The sum rules for $m$ and $\Lambda $ can be found by calculating $\Pi _{\mu
\nu }(p)$ using the parameters of $\mathcal{M}_{\mathrm{AV}}$, as well as by
employing the quark propagators. The function $\Pi _{\mu \nu }^{\mathrm{Phys}%
}(p)$ establishes the phenomenological component of SRs, whereas $\Pi _{\mu
\nu }^{\mathrm{OPE}}(p)$ forms its QCD side.

The correlator $\Pi _{\mu \nu }^{\mathrm{Phys}}(p)$ is given by the
expression
\begin{eqnarray}
\Pi _{\mu \nu }^{\mathrm{Phys}}(p) &=&\frac{\langle 0|J_{\mu }|\mathcal{M}_{%
\mathrm{AV}}(p,\epsilon )\rangle \langle \mathcal{M}_{\mathrm{AV}%
}(p,\epsilon )|J_{\nu }^{\dagger }|0\rangle }{m^{2}-p^{2}}  \notag \\
&&+\cdots .  \label{eq:Phys1}
\end{eqnarray}%
In Eq.\ (\ref{eq:Phys1}) the term presented explicitly is a contribution of
the ground-state particle, whereas effects of higher resonances and
continuum states are shown by ellipses.

To calculate $\Pi _{\mu \nu }^{\mathrm{Phys}}(p)$, it is useful to write it
using the parameters $m$ and $\Lambda $. For these purposes, we introduce
the matrix element
\begin{equation}
\langle 0|J_{\mu }|\mathcal{M}_{\mathrm{AV}}(p,\epsilon )\rangle =\Lambda
\epsilon _{\mu },  \label{eq:ME1}
\end{equation}%
where $\epsilon _{\mu }$ is the polarization vector of the axial-vector
state $\mathcal{M}_{\mathrm{AV}}$. Having performed simple operations, we
find
\begin{equation}
\Pi _{\mu \nu }^{\mathrm{Phys}}(p)=\frac{\Lambda ^{2}}{m^{2}-p^{2}}\left(
-g_{\mu \nu }+\frac{p_{\mu }p_{\nu }}{m^{2}}\right) +\cdots .
\label{eq:Phys2}
\end{equation}%
As is seen, $\Pi _{\mu \nu }^{\mathrm{Phys}}(p)$ is composed of the two
Lorentz structures proportional to $g_{\mu \nu }$ and $p_{\mu }p_{\nu }$.
For our studies, we choose the structure $g_{\mu \nu }$ and denote by $\Pi ^{%
\mathrm{Phys}}(p^{2})$ the corresponding invariant amplitude $\Lambda
^{2}/(m^{2}-p^{2})$.

We calculate $\Pi _{\mu \nu }(p)$ with some accuracy in the operator product
expansion ($\mathrm{OPE}$) by utilizing the heavy quark propagators. The
function $\Pi _{\mu \nu }^{\mathrm{OPE}}(p)$ obtained after this computation
contains the perturbative and nonperturbative terms: the latter is
proportional to $\langle \alpha _{s}G^{2}/\pi \rangle $. The reason is that
heavy quark propagators do not contain light quark and mixed quark-gluon
condensates. As a result, next nonperturbative contributions to $\Pi _{\mu
\nu }^{\mathrm{OPE}}(p)$ are generated by gluon condensates $\langle
g_{s}^{3}G^{3}\rangle $ and $\langle \alpha _{s}G^{2}/\pi \rangle ^{2}$,
which are small and neglected in what follows.

The correlation function $\Pi _{\mu \nu }^{\mathrm{OPE}}(p)$ calculated in
QCD reads
\begin{eqnarray}
&&\Pi _{\mu \nu }^{\mathrm{OPE}}(p)=\frac{i^{2}}{4}\int d^{4}xe^{ipx}\left\{
\mathrm{Tr}\left[ \gamma _{5}S_{c}^{aa^{\prime }}(x)\gamma
_{5}S_{b}^{a^{\prime }a}(-x)\right] \right.  \notag \\
&&\times \mathrm{Tr}\left[ \gamma _{\mu }S_{b}^{bb^{\prime }}(x)\gamma _{\nu
}S_{c}^{b^{\prime }b}(-x)\right] +\mathrm{Tr}\left[ \gamma _{\mu
}S_{c}^{aa^{\prime }}(x)\gamma _{\nu }S_{b}^{a^{\prime }a}(-x)\right]  \notag
\\
&&\left. \times \mathrm{Tr}\left[ \gamma _{5}S_{b}^{bb^{\prime }}(x)\gamma
_{5}S_{c}^{b^{\prime }b}(-x)\right] \right\} ,  \label{eq:QCD1}
\end{eqnarray}%
where $S_{b(c)}(x)$ are propagators of the $b$ and $c$-quarks \cite%
{Agaev:2020zad}, respectively.

The correlator $\Pi _{\mu \nu }^{\mathrm{OPE}}(p)$ also contains two Lorentz
structures mentioned above. To find SRs for the $m$ and $\Lambda $, we
employ the amplitude $\Pi ^{\mathrm{OPE}}(p^{2})$ that corresponds in $\Pi
_{\mu \nu }^{\mathrm{OPE}}(p)$ to the term $g_{\mu \nu }$. Afterwards, by
equating $\Pi ^{\mathrm{OPE}}(p^{2})$ and $\Pi ^{\mathrm{Phys}}(p^{2})$,
applying the Borel transformation and performing continuum subtraction, we
find
\begin{equation}
m^{2}=\frac{\Pi ^{\prime }(M^{2},s_{0})}{\Pi (M^{2},s_{0})},  \label{eq:Mass}
\end{equation}%
and
\begin{equation}
\Lambda ^{2}=e^{m^{2}/M^{2}}\Pi (M^{2},s_{0}),  \label{eq:Coupl}
\end{equation}%
where $\Pi ^{\prime }(M^{2},s_{0})=d\Pi (M^{2},s_{0})/d(-1/M^{2})$. Above $%
\Pi (M^{2},s_{0})$ is the amplitude $\Pi ^{\mathrm{OPE}}(p^{2})$ obtained
after the Borel transformation and continuum subtraction procedures. It is a
function of the Borel and continuum subtraction parameters $M^{2}$ and $%
s_{0} $. The equalities Eqs.\ (\ref{eq:Mass}) and (\ref{eq:Coupl}) are the
sum rules for the mass and current coupling of the molecule $\mathcal{M}_{%
\mathrm{AV}}$.

The amplitude $\Pi (M^{2},s_{0})$ is calculated as an integral of the
spectral density $\rho (s)$%
\begin{equation}
\Pi (M^{2},s_{0})=\int_{4\mathcal{(}m_{b}+m_{c})^{2}}^{s_{0}}ds\rho
(s)e^{-s/M^{2}}+\Pi (M^{2}).
\end{equation}%
The spectral density$\rho (s)$ is equal to the imaginary part of the
invariant amplitude $\Pi ^{\mathrm{OPE}}(p^{2})$ and is a sum of
perturbative $\rho ^{\mathrm{pert.}}(s)$ and nonperturbative $\rho ^{\mathrm{%
Dim4}}(s)$ terms. The function $\Pi (M^{2})$ is the dimension-four
contribution to $\Pi (M^{2},s_{0})$ computed from $\Pi ^{\mathrm{OPE}}(p^{2})
$ and does not contain effects included into $\rho (s)$. The functions $\rho
(s)$ and $\Pi (M^{2})$ are given by the general expression%
\begin{equation}
\rho (s)=\int_{0}^{1}d\alpha \int_{0}^{1-\alpha }d\beta \int_{0}^{1-\alpha
-\beta }d\gamma \rho (s,\alpha ,\beta ,\gamma ),
\end{equation}%
with $\alpha $, $\beta $, and $\gamma $ being the Feynman parameters. The
function $\rho ^{\mathrm{pert.}}(s,\alpha ,\beta ,\gamma )$ has the
following form
\begin{eqnarray}
&&\rho ^{\mathrm{pert.}}(s,\alpha ,\beta ,\gamma )=\frac{3M^{2}\theta (M{)}}{%
4096A^{4}C^{4}\pi ^{6}}[12C^{2}L_{1}^{3}s^{2}\alpha ^{3}\beta ^{3}\gamma ^{3}
\notag \\
&&+4A^{2}Cs\alpha \beta \gamma (3BCm_{b}m_{c}-5L_{1}^{2}M\alpha \beta \gamma
)  \notag \\
&&+3A^{4}(2Cm_{b}m_{c}+M\alpha \beta )(M\gamma L_{1}-2Cm_{b}m_{c})],
\end{eqnarray}%
where $\theta ({z)}$ is the unit step function. Here%
\begin{equation}
M=-C\left[ s\alpha \beta \gamma L_{1}+A[L_{2}m_{c}^{2}-m_{b}^{2}(\beta
+\gamma )]\right] /A^{2},
\end{equation}%
and
\begin{eqnarray}
&&A=\beta \gamma L_{2}+\alpha ^{2}(\beta +\gamma )+\alpha \left[ \beta
^{2}+\gamma (\gamma -1)\right.   \notag \\
&&\left. +\beta (2\gamma -1)\right] ,\ \ B=\alpha ^{2}(\beta -\gamma
)-\gamma L_{2}^{2}  \notag \\
&&+\alpha \lbrack \beta ^{2}+2\gamma (\gamma -1)-\beta (\gamma +1)],  \notag
\\
&&C=\alpha \beta +\alpha \gamma +\beta \gamma .
\end{eqnarray}%
We have also used the notations
\begin{equation}
L_{1}=\alpha +\beta +\gamma -1,\ L_{2}=\beta +\gamma -1.
\end{equation}%
Explicit formulas for $\rho ^{\mathrm{Dim4}}(s,\alpha ,\beta ,\gamma )$ and $%
\Pi (M^{2})$ are rather cumbersome and are not provided here.

To perform numerical analyses it is necessary to fix parameters in Eqs.\ (%
\ref{eq:Mass}) and (\ref{eq:Coupl}). The gluon condensate $\langle \alpha
_{s}G^{2}/\pi \rangle $ and quarks' masses $m_{b}$ and\ $m_{c}$ are
universal quantities: They do not depend on a problem under consideration.
The values of these quantities are listed below
\begin{eqnarray}
&&m_{b}=4.18_{-0.02}^{+0.03}~\mathrm{GeV}\text{,\ }m_{c}=(1.27\pm 0.02)~%
\mathrm{GeV}\text{,}  \notag \\
&&\langle \alpha _{s}G^{2}/\pi \rangle =(0.012\pm 0.004)~\mathrm{GeV}^{4}.
\label{eq:GluonCond}
\end{eqnarray}%
At the same time, the choice of the parameters $M^{2}$ and $s_{0}$ depend on
a task under analysis and has to meet constraints of the SR analysis. In
fact, these parameters should guarantee prevalence of the pole contribution $%
\mathrm{PC}\geq 0.5$ defined as
\begin{equation}
\mathrm{PC}=\frac{\Pi (M^{2},s_{0})}{\Pi (M^{2},\infty )}.  \label{eq:PC}
\end{equation}%
Next restrictions imposed on $M^{2}$ and $s_{0}$ are connected with
convergence of the operator product expansion and stability of extracted $m$
and $\Lambda $ on them. The present computations take into account the
nonperturbative term up to dimension $4$. Contributions of the terms beyond $%
\langle \alpha _{s}G^{2}/\pi \rangle $, i.e., ones proportional to dimension
six and eight condensates $\langle g_{s}^{3}G^{3}\rangle $ and $\langle \alpha
_{s}G^{2}/\pi \rangle ^{2}$ are negligibly small here. Indeed, as we shall
just see below in the considered region of the parameters $M^{2}$ and $s_{0}$
even the dimension-four contribution forms only $1\%$ of the full result.
Therefore to ensure convergence of $\mathrm{OPE}$ we impose a constraint on
this term and require fulfilment of the condition
\begin{equation}
|\Pi ^{\mathrm{Dim4}}(M^{2},s_{0})|\leq 0.05|\Pi (M^{2},s_{0})|.
\label{eq:OPE}
\end{equation}%
\begin{figure}[h]
\includegraphics[width=8.5cm]{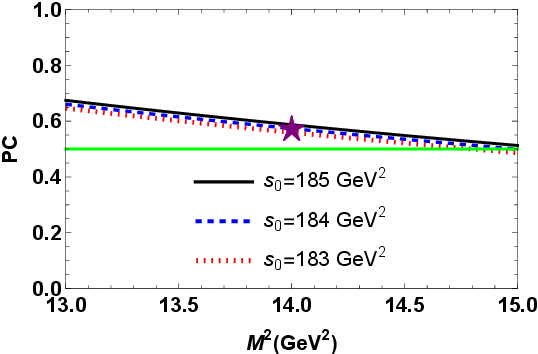}
\caption{The pole contribution $\mathrm{PC}$ as a function of $M^{2}$ at
fixed $s_{0}$. The red star shows the point $M^{2}=14~\mathrm{GeV}^{2}$ and $%
s_{0}=184~\mathrm{GeV}^{2}$. }
\label{fig:PC}
\end{figure}

Our calculations demonstrate that the working regions
\begin{equation}
M^{2}\in \lbrack 13,15]~\mathrm{GeV}^{2},\ s_{0}\in \lbrack 183,185]~\mathrm{%
GeV}^{2},  \label{eq:Wind1}
\end{equation}%
meet these constraints. In fact, at $M^{2}=15~\mathrm{GeV}^{2}$ we find $%
\mathrm{PC}\approx 0.50$ (on the average in $s_{0})$, whereas at $M^{2}=13~%
\mathrm{GeV}^{2}$ the pole contribution is equal to $\mathrm{PC}\approx 0.66$%
. The term $|\Pi ^{\mathrm{Dim4}}(M^{2},s_{0})|$ at $M^{2}=13~\mathrm{GeV}%
^{2}$ does not exceed $1\%$ of the whole result. In Fig.\ \ref{fig:PC}, we
plot dependence of $\mathrm{PC}$ on the parameter $M^{2}$ at some values of $%
s_{0}$. It is clear, that\ $\mathrm{PC}$ overshoots $0.5$ for all $M^{2}$
and $s_{0}$ from the regions Eq.\ (\ref{eq:Wind1}).

\begin{widetext}

\begin{figure}[h!]
\begin{center}
\includegraphics[totalheight=6cm,width=8cm]{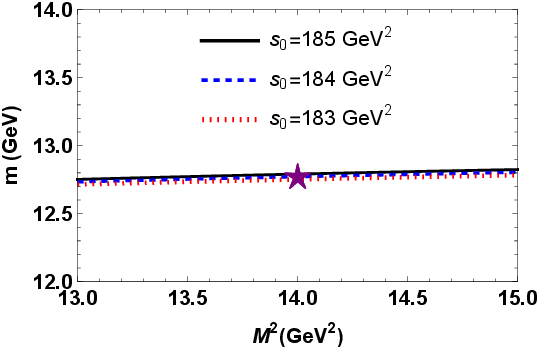}
\includegraphics[totalheight=6cm,width=8cm]{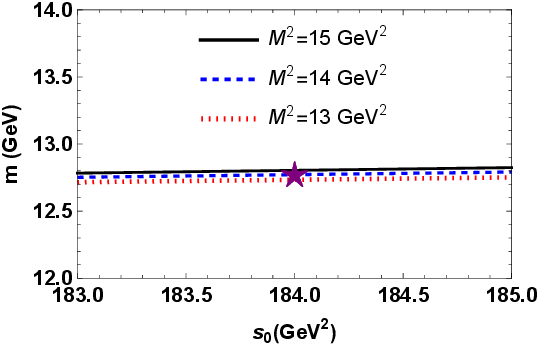}
\end{center}
\caption{Mass $m$ of the hadronic molecule $\mathcal{M}_{\mathrm{AV}}$ vs the parameters $M^{2}$ (left panel), and $s_0$ (right panel).}
\label{fig:Mass}
\end{figure}

\end{widetext}

The mass and current coupling of the molecule $\mathcal{M}_{\mathrm{AV}}$
are evaluated as their average values in domains Eq.\ (\ref{eq:Wind1}), and
are equal to
\begin{eqnarray}
m &=&(12770\pm 60)~\mathrm{MeV},  \notag \\
\Lambda &=&(1.03\pm 0.06)~\mathrm{GeV}^{5}.  \label{eq:Result1}
\end{eqnarray}%
These results coincide with the sum rule predictions at the point $M^{2}=14~%
\mathrm{GeV}^{2}$ and $s_{0}=184~\mathrm{GeV}^{2}$, where $\mathrm{PC}%
\approx 0.57$. This guarantee the dominance of $\mathrm{PC}$ in the
extracted results, and confirms the ground-state nature of the molecule $%
\mathcal{M}_{\mathrm{AV}}$ in its class. The mass $ m $ as a function of the parameters $ M^2 $ and $ s_0 $ is shown in Fig.  \ref{fig:Mass}.

%%%%%%%%%%%%%%%%%%%%%%%%%%%%%%%%%%%%%%%%%%%%%%%%%%%%%%%%%%%%%%%%%

\section{Fall-apart decays of $\mathcal{M}_{\mathrm{AV}}$}

\label{sec:Widths1}

%%%%%%%%%%%%%%%%%%%%%%%%%%%%%%%%%%%%%%%%%%%%%%%%%%%%%%%%%%%

Here, we explore dominant decays of the axial-vector molecule $\mathcal{M}_{%
\mathrm{AV}}$ into ordinary mesons and calculate partial widths of these
processes. These decays are so-called fall-apart channels in which all
valence quarks from $\mathcal{M}_{\mathrm{AV}}$ participate in forming of
the final-state mesons. We rely on the mass $m$ and spin-parity $J^{\mathrm{%
PC}}=1^{++}$ of the molecule $\mathcal{M}_{\mathrm{AV}}$ to determine its
allowed decay channels. It is not difficult to see that decays into pairs of
$J/\psi \eta _{b}$, $\Upsilon \eta _{c}$, $B_{c}^{\ast +}B_{c}^{-}$ and $%
B_{c}^{\ast -}B_{c}^{+}$ mesons are permitted channels for transforming of $%
\mathcal{M}_{\mathrm{AV}}$ to conventional particles.

%%%%%%%%%%%%%%%%%%%%%%%%%%%%%%%%%%%%%%%%%%%%%%%%%%%%%%%%%%%%%%%%%%5

\subsection{$\mathcal{M}_{\mathrm{AV}}\rightarrow J/\protect\psi \protect%
\eta _{b}$}

%%%%%%%%%%%%%%%%%%%%%%%%%%%%%%%%%%%%%%%%%%%%%%%%%%%%%%%%%%%%%%%%%%%%%%%

Firstly, we consider the decay $\mathcal{M}_{\mathrm{AV}}\rightarrow J/\psi
\eta _{b}$. The width of this process depends on the numerous parameters
including the masses, decay constants, current couplings of the molecule and
final-state mesons. But, it also contains the strong coupling $g_{1}$ of
particles at the vertex $\mathcal{M}_{\mathrm{AV}}J/\psi \eta _{b}$. Our aim
is to calculate the coupling $g_{1}$.

To this end, we examine the three-point QCD correlator%
\begin{eqnarray}
\Pi _{1\mu \nu }(p,p^{\prime }) &=&i^{2}\int d^{4}xd^{4}ye^{ip^{\prime
}y}e^{-ipx}\langle 0|\mathcal{T}\{J^{\eta _{b}}(y)  \notag \\
&&\times J_{\mu }^{J/\psi }(0)J_{\nu }^{\dagger }(x)\}|0\rangle ,
\label{eq:CF3}
\end{eqnarray}%
with
\begin{equation}
J^{\eta _{b}}(x)=\overline{b}_{i}(x)i\gamma _{5}b_{i}(x),\ J_{\mu }^{J/\psi
}(x)=\overline{c}_{j}(x)\gamma _{\mu }c_{j}(x),  \label{eq:CR3}
\end{equation}%
being the $\eta _{b}$ and $J/\psi $ mesons' interpolating currents,
respectively.

Analysis of this correlation function permits one to find the SR for the
form factor $g_{1}(q^{2})$, which at the mass shell $q^{2}=m_{J/\psi }^{2}$
gives $g_{1}$. To derive the required sum rule, we recast $\Pi _{1\mu \nu
}(p,p^{\prime })$ into the form%
\begin{eqnarray}
&&\Pi _{1\mu \nu }^{\mathrm{Phys}}(p,p^{\prime })=\frac{\langle 0|J^{\eta
_{b}}|\eta _{b}(p^{\prime })\rangle }{p^{\prime 2}-m_{\eta _{b}}^{2}}\frac{%
\langle 0|J_{\mu }^{J/\psi }|J/\psi (q,\varepsilon )\rangle }{%
q^{2}-m_{J/\psi }^{2}}  \notag \\
&&\times \langle \eta _{b}(p^{\prime })J_{\mu }^{J/\psi }(q,\varepsilon )|%
\mathcal{M}_{\mathrm{AV}}(p,\epsilon )\rangle \frac{\langle \mathcal{M}_{%
\mathrm{AV}}(p,\epsilon )|J_{\nu }^{\dagger }|0\rangle }{p^{2}-m^{2}}+\cdots
,  \notag \\
&&  \label{eq:CF5}
\end{eqnarray}%
where $m_{\eta _{b}}=(9398.7\pm 2.0)~\mathrm{MeV}$ and $m_{J/\psi
}=(3096.900\pm 0.006)~\mathrm{MeV}$ are the masses of $\eta _{b}$ and $%
J/\psi $ \cite{PDG:2024}, respectively. The correlator $\Pi _{1\mu \nu }^{%
\mathrm{Phys}}(p,p^{\prime })$ is a contribution of the ground-state
particles, whereas the dots denote effects of higher resonances and
continuum states. Next, we introduce the matrix elements
\begin{eqnarray}
\langle 0|J^{\eta _{b}}|\eta _{b}(p^{\prime })\rangle &=&\frac{f_{\eta
_{b}}m_{\eta _{b}}^{2}}{2m_{b}},\   \notag \\
\langle 0|J_{\mu }^{J/\psi }|J/\psi (q,\varepsilon )\rangle &=&f_{J/\psi
}m_{J/\psi }\varepsilon _{\mu }(q).  \label{eq:ME2}
\end{eqnarray}%
where $f_{\eta _{b}}=724~\mathrm{MeV}$ and $f_{J/\psi }=(411\pm 7)~\mathrm{%
MeV}$ are decay constants of the corresponding mesons.

The vertex $\mathcal{M}_{\mathrm{AV}}J/\psi \eta _{b}$ is given by the
formula
\begin{eqnarray}
&&\langle \eta _{b}(p^{\prime })J/\psi (q,\varepsilon )|\mathcal{M}_{\mathrm{%
AV}}(p,\epsilon )\rangle =g_{1}(q^{2})  \notag \\
&&\times \lbrack p\cdot p^{\prime }\epsilon \cdot \varepsilon ^{\ast
}-p\cdot \varepsilon ^{\ast }p^{\prime }\cdot \epsilon ].  \label{eq:ME3}
\end{eqnarray}%
After some computations, for $\Pi _{1\mu \nu }^{\mathrm{Phys}}(p,p^{\prime
}) $ one obtains
\begin{eqnarray}
&&\Pi _{1\mu \nu }^{\mathrm{Phys}}(p,p^{\prime })=g_{1}(q^{2})\frac{\Lambda
f_{\eta _{b}}m_{\eta _{b}}^{2}f_{J/\psi }m_{J/\psi }}{2m_{b}\left(
p^{2}-m^{2}\right) \left( p^{\prime 2}-m_{\eta _{b}}^{2}\right) }  \notag \\
&&\times \frac{1}{(q^{2}-m_{J/\psi }^{2})}\left[ \frac{m^{2}+m_{\eta
_{b}}^{2}-q^{2}}{2}g_{\mu \nu }-\frac{m^{2}}{m_{J/\psi }^{2}}p_{\mu
}^{\prime }p_{\nu }^{\prime }\right.  \notag \\
&&\left. -\frac{m^{2}+m_{\eta _{b}}^{2}-q^{2}}{2m_{J/\psi }^{2}}(p_{\mu
}p_{\nu }-p_{\mu }p_{\nu }^{\prime })+\frac{m^{2}-m_{J/\psi }^{2}}{m_{J/\psi
}^{2}}p_{\mu }^{\prime }p_{\nu }\right]  \notag \\
&&+\cdots .  \label{eq:CF6}
\end{eqnarray}%
As is seen, the correlator $\Pi _{1\mu \nu }^{\mathrm{Phys}}(p,p^{\prime })$
has rather different Lorentz structures, therefore we choose, for our
purposes, the term $\sim g_{\mu \nu }$ and denote the corresponding
invariant amplitude by $\Pi _{1}^{\mathrm{Phys}}(p^{2},p^{\prime 2},q^{2})$.

The same correlation function $\Pi _{1\mu \nu }(p,p^{\prime })$ expressed
using the heavy quark propagators reads
\begin{eqnarray}
&&\Pi _{1\mu \nu }^{\mathrm{OPE}}(p,p^{\prime })=\frac{1}{2}\int
d^{4}xd^{4}ye^{ip^{\prime }y}e^{-ipx}\left\{ \mathrm{Tr}\left[ \gamma
_{5}S_{b}^{ib}(y-x)\right. \right.  \notag \\
&&\left. \times \gamma _{\nu }S_{c}^{bj}(x)\gamma _{\mu
}S_{c}^{ja}(-x)\gamma _{5}S_{b}^{ai}(x-y)\right] -\mathrm{Tr}\left[ \gamma
_{5}S_{b}^{ib}(y-x)\right.  \notag \\
&&\left. \left. \times \gamma _{5}S_{c}^{bj}(x)\gamma _{\mu
}S_{c}^{ja}(-x)\gamma _{\nu }S_{b}^{ai}(x-y)\right] \right\} .
\label{eq:QCDside2}
\end{eqnarray}%
The correlator $\Pi _{1\mu \nu }^{\mathrm{OPE}}(p,p^{\prime })$ has the
terms with the same Lorentz structures. We label by $\Pi _{1}^{\mathrm{OPE}%
}(p^{2},p^{\prime 2},q^{2})$ the amplitude that corresponds to term $\sim
g_{\mu \nu }$ in the correlator $\Pi _{1\mu \nu }^{\mathrm{OPE}}(p,p^{\prime
})$. This amplitude after the double Borel transformations over $-p^{2}$, $%
-p^{\prime 2}$ and continuum subtractions determines the SR for $%
g_{1}(q^{2}) $%
\begin{eqnarray}
&&g_{1}(q^{2})=\frac{4m_{b}}{\Lambda f_{\eta _{b}}m_{\eta _{b}}^{2}f_{J/\psi
}m_{J/\psi }}\frac{q^{2}-m_{J/\psi }^{2}}{m^{2}+m_{\eta _{b}}^{2}-q^{2}}
\notag \\
&&\times e^{m^{2}/M_{1}^{2}}e^{m_{\eta _{b}}^{2}/M_{2}^{2}}\Pi _{1}(\mathbf{M%
}^{2},\mathbf{s}_{0},q^{2}).  \label{eq:SRCoup2}
\end{eqnarray}%
In Eq.\ (\ref{eq:SRCoup2}) $\Pi _{1}(\mathbf{M}^{2},\mathbf{s}_{0},q^{2})$
is the Borel transformed and subtracted function $\Pi _{1}^{\mathrm{OPE}%
}(p^{2},p^{\prime 2},q^{2})$. It depends on the parameters $\mathbf{M}%
^{2}=(M_{1}^{2},M_{2}^{2})$ and $\mathbf{s}_{0}=(s_{0},s_{0}^{\prime })$
where the pairs $(M_{1}^{2},s_{0})$ and $(M_{2}^{2},s_{0}^{\prime })$
correspond to the hadronic molecule and $\eta _{b}$ channels, respectively.

The function $\Pi _{1}(\mathbf{M}^{2},\mathbf{s}_{0},q^{2})$ is given by the
expression%
\begin{eqnarray}
&&\Pi _{1}(\mathbf{M}^{2},\mathbf{s}_{0},q^{2})=%
\int_{4(m_{b}+m_{c})^{2}}^{s_{0}}ds\int_{4m_{b}^{2}}^{s_{0}^{\prime
}}ds^{\prime }\rho _{1}(s,s^{\prime },q^{2})  \notag \\
&&\times e^{-s/M_{1}^{2}}e^{-s^{\prime }/M_{2}^{2}}.  \label{eq:SRG}
\end{eqnarray}%
The spectral density $\rho _{1}(s,s^{\prime },q^{2})$ is evaluated as the
imaginary part of the amplitude $\Pi _{1}(p^{2},p^{\prime 2},q^{2})$, and
has the following form
\begin{eqnarray}
&&\rho _{1}(s,s^{\prime },q^{2})=\frac{1}{128\pi ^{4}}\int_{0}^{1}d\alpha
\int_{0}^{1-\alpha }d\beta \int_{0}^{1-\alpha -\beta }d\gamma  \notag \\
&&\times \frac{\theta (N)F}{(\alpha +\gamma )^{9}(\alpha +\gamma -1)^{4}},
\label{eq:Rho1}
\end{eqnarray}%
where
\begin{eqnarray}
&&F=9m_{c}m_{b}(s-s^{\prime }-q^{2})(\alpha +\gamma )^{3}(\alpha +\gamma
-1)^{2}  \notag \\
&&+12[m_{c}^{2}(\alpha +\gamma )^{2}+q^{2}\alpha \gamma ]\left[ -3s^{\prime
}\beta L_{1}(\alpha +\gamma )\right.  \notag \\
&&\left. +2(\alpha +\gamma -1)(m_{c}^{2}(\alpha +\gamma )^{2}-q^{2}\alpha
\gamma )\right] .  \label{eq:Dec1}
\end{eqnarray}%
Here, the unit step function $\theta (N)$ has the argument
\begin{eqnarray}
&&N=\frac{1}{(\alpha +\gamma -1)^{2}(\alpha +\gamma )}\left\{ \left[
m_{c}^{2}(\alpha +\gamma )(\alpha +\gamma -1)\right. \right.  \notag \\
&&\left. \left. -s^{\prime }\beta L_{1}\right] (\alpha +\gamma )-q^{2}\alpha
\gamma (\alpha +\gamma )\right\} -m_{b}^{2}.
\end{eqnarray}

To perform numerical analysis, we employ in the $\mathcal{M}_{\mathrm{AV}}$
channel $M_{1}^{2}$ and $s_{0}$ from Eq.\ (\ref{eq:Wind1}). In the $\eta
_{b} $ channel,\ we use as the parameters $(M_{2}^{2},\ s_{0}^{\prime })$
\begin{equation}
M_{2}^{2}\in \lbrack 9,11]~\mathrm{GeV}^{2},\ s_{0}^{\prime }\in \lbrack
95,99]~\mathrm{GeV}^{2},  \label{eq:Wind3}
\end{equation}%
where the upper limit $\sqrt{s_{0}^{\prime }}$ is restricted by the mass $%
9.999\ \mathrm{GeV}$ of the first radially excited $\eta _{b}(2S)$ meson.

The credible results for the form factor $g_{1}(q^{2})$ in the sum rule
context is obtained at $q^{2}<0$. But the coupling $g_{1}$ should be
determined at $q^{2}=m_{J/\psi }^{2}$. This problem can be avoided by
introducing a variable $Q^{2}=-q^{2}$ and employing the notation $%
g_{1}(Q^{2})$ for the new function. Afterwards, we use fitting functions
which for $Q^{2}>0$ coincides with the SR data, but can also be applied at $%
Q^{2}<0$.

In general there are different functions to extrapolate the QCD data to the
region of negative $Q^{2}$. To this end, we introduce the fit functions $%
\mathcal{G}_{1}(Q^{2})$ and $\overline{\mathcal{G}}_{1}(Q^{2})$%
\begin{equation}
\mathcal{G}_{i}(Q^{2})=\mathcal{G}_{i}^{0}\mathrm{\exp }\left[ a_{i}^{1}%
\frac{Q^{2}}{m^{2}}+a_{i}^{2}\left( \frac{Q^{2}}{m^{2}}\right) ^{2}\right] ,
\label{eq:FitF}
\end{equation}%
and%
\begin{equation}
\overline{\mathcal{G}}_{i}(Q^{2})=\frac{\overline{\mathcal{G}}_{i}^{0}}{%
\left( 1-\frac{Q^{2}}{m^{2}}\right) \left( 1-\sigma _{i}^{1}\frac{Q^{2}}{%
m^{2}}+\sigma _{i}^{2}\left( \frac{Q^{2}}{m^{2}}\right) ^{2}\right) },
\label{eq:FitF2}
\end{equation}%
where $\mathcal{G}_{i}^{0}$, $a_{i}^{1}$, and $a_{i}^{2},$ as well as $%
\overline{\mathcal{G}}_{i}^{0}$, $\sigma _{i}^{1}$ and $\sigma _{i}^{2}$ are
parameters extracted from comparison with the SR data.

\begin{figure}[h]
\includegraphics[width=8.5cm]{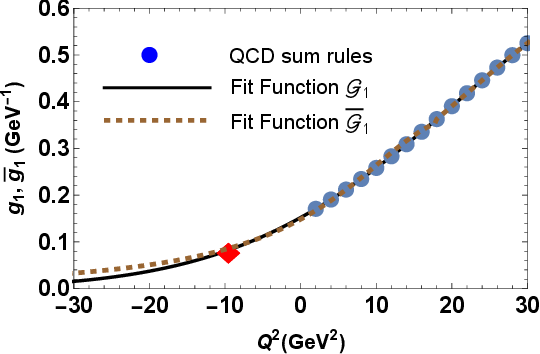}
\caption{The sum rule's data and extrapolating functions for the form
factors $g_{1}(Q^{2})$ and $\overline{g}_{1}(Q^{2})$. The red diamond fixes
the point $Q^{2}=-m_{J/\protect\psi }^{2}$. }
\label{fig:Fit}
\end{figure}

Then, from analysis of the QCD data and Eqs.\ (\ref{eq:FitF}) and (\ref%
{eq:FitF2}), we find the parameters $\mathcal{G}_{1}^{0}=0.15~\mathrm{GeV}%
^{-1}$, $a_{1}^{1}=9.58$, and $a_{1}^{2}=-15.47$ of the function $\mathcal{G}%
_{1}(Q^{2})$, and $\overline{\mathcal{G}}_{i}^{0}=0.15~\mathrm{GeV}^{-1}$, $%
\sigma _{1}^{1}=8.72$ and $\sigma _{1}^{2}=35.46$ of the function $\overline{%
\mathcal{G}}_{1}(Q^{2})$. They are shown in Fig.\ \ref{fig:Fit}, in which
one sees very nice agreement of $\mathcal{G}_{1}(Q^{2})$ and SR data. Having
used this function, we get $g_{1}$
\begin{equation}
g_{1}\equiv \mathcal{G}_{1}(-m_{J/\psi }^{2})=(8.20\pm 1.48)\times 10^{-2}\
\mathrm{GeV}^{-1}.  \label{eq:g1}
\end{equation}%
Agreement of the function $\overline{\mathcal{G}}_{1}(Q^{2})$ with SR data
is nice in the region of interest. At $Q^{2}=-m_{J/\psi }^{2}$ we find
\begin{equation}
\overline{g}_{1}\equiv \overline{\mathcal{G}}_{1}(-m_{J/\psi }^{2})=(8.57\pm
1.51)\times 10^{-2}\ \mathrm{GeV}^{-1}
\end{equation}%
that is very close to $g_{1}$. The difference $0.0037$ between these two
predictions is smaller than the corresponding theoretical errors. Therefore,
throughout this work we employ Eq.\ (\ref{eq:FitF}) and neglect small
effects due to alternative extrapolating functions.

We calculate the width of the decay $\mathcal{M}_{\mathrm{AV}}\rightarrow
J/\psi \eta _{b}$ by means of the formula%
\begin{equation}
\Gamma \left[ \mathcal{M}_{\mathrm{AV}}\rightarrow J/\psi \eta _{b}\right]
=g_{1}^{2}\frac{\lambda _{1}}{24\pi }|M_{1}|^{2},
\end{equation}%
where
\begin{eqnarray}
|M_{1}|^{2} &=&\frac{1}{4m_{J/\psi }^{2}}\left[ m^{6}-2m^{4}m_{\eta
_{b}}^{2}+2(m_{J/\psi }^{3}-m_{\eta _{b}}^{2}m_{J/\psi })^{2}\right.  \notag
\\
&&\left. +m^{2}(m_{\eta _{b}}^{4}+6m_{\eta _{b}}^{2}m_{J/\psi
}^{2}-3m_{J/\psi }^{4})\right] .  \label{eq:PDw2}
\end{eqnarray}%
Here, $\lambda _{1}=\lambda (m,m_{\eta _{b}},m_{J/\psi })$ is given by the
expression
\begin{equation}
\lambda (x,y,z)=\frac{\sqrt{%
x^{4}+y^{4}+z^{4}-2(x^{2}y^{2}+x^{2}z^{2}+y^{2}z^{2})}}{2x}.
\end{equation}%
As a result, we obtain
\begin{equation}
\Gamma \left[ \mathcal{M}_{\mathrm{AV}}\rightarrow J/\psi \eta _{b}\right]
=(29.6\pm 8.2)~\mathrm{MeV}.  \label{eq:DW1}
\end{equation}

A part $\pm 7.6$ of the total errors in Eq.\ (\ref{eq:DW1}) is connected
with the uncertainties of the coupling $g_{1}$, whereas $\pm 3.0$ is
generated by the ambiguities of the particles' masses in Eq.\ (\ref{eq:PDw2}%
).

%%%%%%%%%%%%%%%%%%%%%%%%%%%%%%%%%%%%%%%%%%%%%%%%%%%%%%%%%%%%%%%%%%5

\subsection{$\mathcal{M}_{\mathrm{AV}}\rightarrow \Upsilon \protect\eta _{c}$%
}

%%%%%%%%%%%%%%%%%%%%%%%%%%%%%%%%%%%%%%%%%%%%%%%%%%%%%%%%%%%%%%%%%%%%%%%

Here, we consider the decay $\mathcal{M}_{\mathrm{AV}}\rightarrow \Upsilon
\eta _{c}$ and determine the strong coupling $g_{2}$ at the vertex $\mathcal{%
M}_{\mathrm{AV}}\Upsilon \eta _{c}$ which amounts to the form factor $%
g_{2}(q^{2})$ at the mass shell $q^{2}=m_{\eta _{c}}^{2}$.

The SR for the form factor $g_{2}(q^{2})$ is derived by means of the
correlation function
\begin{eqnarray}
\Pi _{2\mu \nu }(p,p^{\prime }) &=&i^{2}\int d^{4}xd^{4}ye^{ip^{\prime
}y}e^{-ipx}\langle 0|\mathcal{T}\{J_{\mu }^{\Upsilon }(y)  \notag \\
&&\times J^{\eta _{c}}(0)J_{\nu }^{\dagger }(x)\}|0\rangle ,  \label{eq:CF1a}
\end{eqnarray}%
where $J_{\mu }^{\Upsilon }(x)$ and $J^{\eta _{c}}(x)$ are the interpolating
currents of the $\Upsilon $ and $\eta _{c}$ mesons, respectively. They are
defined as
\begin{equation}
J_{\mu }^{\Upsilon }(x)=\overline{b}_{i}(x)\gamma _{\mu }b_{i}(x),~J^{\eta
_{c}}(x)=\overline{c}_{j}(x)i\gamma _{5}c_{j}(x),
\end{equation}%
with $i$ and $j$ being the color indices.

To find the phenomenological component $\Pi _{2\mu \nu }^{\mathrm{Phys}%
}(p,p^{\prime })$ of the sum rule, we need to rewrite Eq.\ (\ref{eq:CF1a})
using the involved particles' physical parameters. By taking into account
only contributions of the ground-state particles, we recast the correlator $%
\Pi _{2\mu \nu }(p,p^{\prime })$ into the form%
\begin{eqnarray}
&&\Pi _{2\mu \nu }^{\mathrm{Phys}}(p,p^{\prime })=g_{2}(q^{2})\frac{\Lambda
f_{\eta _{c}}m_{\eta _{c}}^{2}f_{\Upsilon }m_{\Upsilon }}{2m_{c}\left(
p^{2}-m^{2}\right) \left( p^{\prime 2}-m_{\Upsilon }^{2}\right) }  \notag \\
&&\times \frac{1}{(q^{2}-m_{\eta _{c}}^{2})}\left[ \frac{m^{2}-m_{\Upsilon
}^{2}+q^{2}}{2}g_{\mu \nu }-p_{\mu }p_{\nu }\right.  \notag \\
&&\left. +p_{\mu }^{\prime }p_{\nu }-\frac{m^{2}}{m_{\Upsilon }^{2}}p_{\mu
}^{\prime }p_{\nu }^{\prime }+\frac{m^{2}+m_{\Upsilon }^{2}-q^{2}}{%
2m_{\Upsilon }^{2}}p_{\mu }p_{\nu }^{\prime }\right]  \notag \\
&&+\cdots .  \label{eq:Phys3}
\end{eqnarray}%
where $m_{\eta _{c}}=(2984.1\pm 0.4)~\mathrm{MeV}$ and $m_{\Upsilon
}=(9460.40\pm0.10)~\mathrm{MeV}\ $ are the masses of the mesons PDG \cite%
{PDG:2024}. The $f_{\eta _{c}}=(421\pm 35)~\mathrm{MeV}$ and $f_{\Upsilon
}=(708\pm 8)~\mathrm{MeV}$ are their decay constants: For $f_{\eta _{c}}$ we
use the SR prediction \cite{Veliev:2010vd}, whereas for $f_{\Upsilon }$
employ its experimental value borrowed from Ref.\ \cite{Lakhina:2006vg}. In
deriving Eq.\ (\ref{eq:Phys3}), we have used the following matrix elements

\begin{eqnarray}
&&\langle 0|J_{\mu }^{\Upsilon }|\Upsilon (p^{\prime },\varepsilon )\rangle
=f_{\Upsilon }m_{\Upsilon }\varepsilon _{\mu }(p^{\prime }),  \notag \\
&&\langle 0|J^{\eta _{c}}|\eta _{c}(q)\rangle =\frac{f_{\eta _{c}}m_{\eta
_{c}}^{2}}{2m_{c}},  \label{eq:ME4}
\end{eqnarray}%
and%
\begin{eqnarray}
&&\langle \eta _{c}(q)\Upsilon (p^{\prime },\varepsilon )|\mathcal{M}_{%
\mathrm{AV}}(p,\epsilon )\rangle =g_{2}(q^{2})\left[ p\cdot q\right.  \notag
\\
&&\left. \times \epsilon \cdot \varepsilon ^{\ast }-p\cdot \varepsilon
^{\ast }q\cdot \epsilon \right] .
\end{eqnarray}

The QCD side of the sum rule $\Pi _{2\mu \nu }^{\mathrm{OPE}}(p,p^{\prime })$
is determined by the formula
\begin{eqnarray}
&&\Pi _{2\mu \nu }^{\mathrm{OPE}}(p,p^{\prime })=\frac{1}{2}\int
d^{4}xd^{4}ye^{ip^{\prime }y}e^{-ipx}\left\{ \mathrm{Tr}\left[ \gamma _{\mu
}S_{b}^{ib}(y-x)\right. \right.  \notag \\
&&\left. \times \gamma _{\nu }S_{c}^{bj}(x)\gamma _{5}S_{c}^{ja}(-x)\gamma
_{5}S_{b}^{ai}(x-y)\right] -\mathrm{Tr}\left[ \gamma _{\mu
}S_{b}^{ib}(y-x)\right.  \notag \\
&&\left. \left. \times \gamma _{5}S_{c}^{bj}(x)\gamma
_{5}S_{c}^{ja}(-x)\gamma _{\nu }S_{b}^{ai}(x-y)\right] \right\} .
\end{eqnarray}%
The functions $\Pi _{2\mu \nu }^{\mathrm{Phys}}(p,p^{\prime })$ and $\Pi
_{2\mu \nu }^{\mathrm{OPE}}(p,p^{\prime })$ contain the similar terms. Here,
we pick up components $\sim g_{\mu \nu }$ and employ relevant invariant
amplitudes to get the sum rule for $g_{2}(q^{2})$

\begin{eqnarray}
&&g_{2}(q^{2})=\frac{4m_{c}}{\Lambda f_{\eta _{c}}m_{\eta
_{c}}^{2}f_{\Upsilon }m_{\Upsilon }}\frac{q^{2}-m_{\eta _{c}}^{2}}{%
m^{2}-m_{\Upsilon }^{2}+q^{2}}  \notag \\
&&\times e^{m^{2}/M_{1}^{2}}e^{m_{\Upsilon }^{2}/M_{2}^{2}}\Pi _{2}(\mathbf{M%
}^{2},\mathbf{s}_{0},q^{2}),
\end{eqnarray}%
where $\Pi _{2}(\mathbf{M}^{2},\mathbf{s}_{0},q^{2})$ is Borel transformed
and subtracted amplitude $\Pi _{2}^{\mathrm{OPE}}(p^{2},p^{\prime 2},q^{2})$.

Operations required to carry out numerical analysis are ordinary
manipulations explained above. Therefore, we omit further details and
provide necessary final information. Thus, for the parameters $M_{2}^{2}$,
and $s_{0}^{\prime }$ in the $\Upsilon $ channel, we have used the regions
\begin{equation}
M_{2}^{2}\in \lbrack 10,12]~\mathrm{GeV}^{2},\ s_{0}^{\prime }\in \lbrack
98,100]~\mathrm{GeV}^{2}.
\end{equation}%
Calculations of $g_{2}(Q^{2})$ is done in the domain $Q^{2}=2-30~\mathrm{GeV}%
^{2}$. The extrapolating function $\mathcal{G}_{2}(Q^{2})$ has the
parameters $\mathcal{G}_{2}^{0}=0.46~\mathrm{GeV}^{-1}$, $a_{2}^{1}=12.04$,
and $a_{2}^{2}=-8.81$. As a result, the strong coupling $g_{2}$ is
\begin{equation}
g_{2}\equiv \mathcal{G}_{2}(-m_{\eta _{c}}^{2})=(2.3\pm 0.4)\times 10^{-1}\
\mathrm{GeV}^{-1}.
\end{equation}%
Then, for the width of the decay $\mathcal{M}_{\mathrm{AV}}\rightarrow
\Upsilon \eta _{c}$ we obtain
\begin{equation}
\Gamma \left[ \mathcal{M}_{\mathrm{AV}}\rightarrow \Upsilon \eta _{c}\right]
=(28.7\pm 7.9)~\mathrm{MeV}.  \label{eq:PW2}
\end{equation}%
The errors in Eq.\ (\ref{eq:PW2}) are connected with ambiguities of the
coupling $g_{2}$ and masses $m$, $m_{\Upsilon }$ and $m_{\eta _{c}}$.

.

%%%%%%%%%%%%%%%%%%%%%%%%%%%%%%%%%%%%%%%%%%%%%%%%%%%%%%%%%%%%%%%%%%5

\subsection{$\mathcal{M}_{\mathrm{AV}}\rightarrow B_{c}^{\ast +}B_{c}^{-}$}

%%%%%%%%%%%%%%%%%%%%%%%%%%%%%%%%%%%%%%%%%%%%%%%%%%%%%%%%%%%%%%%%%%%%%%%

The partial width of the decay $\mathcal{M}_{\mathrm{AV}}\rightarrow
B_{c}^{\ast +}B_{c}^{-}$ is determined by the coupling $g_{3}$ at the vertex
$\mathcal{M}_{\mathrm{AV}}B_{c}^{\ast +}B_{c}^{-}$. In the framework of the
QCD SR method the form factor $g_{3}(q^{2})$ is calculated using the
three-point correlation function%
\begin{eqnarray}
&&\Pi _{3\mu \nu }(p,p^{\prime })=i^{2}\int d^{4}xd^{4}ye^{ip^{\prime
}y}e^{-ipx}\langle 0|\mathcal{T}\{J_{\mu }^{B_{c}^{\ast }}(y)  \notag \\
&&\times J^{B_{c}}(0)J_{\nu }^{\dagger }(x)\}|0\rangle .  \label{eq:CF7}
\end{eqnarray}%
In the function above $J_{\mu }^{B_{c}^{\ast }}(x)$ and $J^{B_{c}}(x)$ are
the interpolating currents for the mesons $B_{c}^{\ast +}$ and $B_{c}^{-}$,
respectively
\begin{equation}
J_{\mu }^{B_{c}^{\ast }}(x)=\overline{b}_{i}(x)\gamma _{\mu }c_{i}(x),\
J^{B_{c}}(x)=\overline{c}_{j}(x)i\gamma _{5}b_{j}(x).
\end{equation}

The correlator $\Pi _{3\mu \nu }(p,p^{\prime })$ in terms of the involved
particles' matrix elements reads

\begin{eqnarray}
&&\Pi _{3\mu \nu }^{\mathrm{Phys}}(p,p^{\prime })=\frac{\langle 0|J_{\nu
}^{B_{c}^{\ast }}|B_{c}^{\ast +}(p^{\prime },\varepsilon )\rangle }{%
p^{\prime 2}-m_{B_{c}^{\ast }}^{2}}\frac{\langle
0|J^{B_{c}}|B_{c}^{-}(q)\rangle }{q^{2}-m_{B_{c}}^{2}}  \notag \\
&&\times \langle B_{c}^{\ast +}(p^{\prime },\varepsilon )B_{c}^{-}(q)|%
\mathcal{M}_{\mathrm{AV}}(p,\epsilon )\rangle \frac{\langle \mathcal{M}_{%
\mathrm{AV}}(p,\epsilon )|J^{\dagger }|0\rangle }{p^{2}-m^{2}}  \notag \\
&&+\cdots ,
\end{eqnarray}%
where $m_{B_{c}}=6274.47(27)(17)~\mathrm{MeV}$ and $m_{B_{c}^{\ast }}=6338~%
\mathrm{MeV}$ are the masses of the mesons $B_{c}^{\ast +}$ and $B_{c}^{-}$,
respectively. As the mass of the $B_{c}$ meson we use its experimental
value, whereas $m_{B_{c}^{\ast }}$ is theoretical prediction from Ref.\ \cite%
{Godfrey:2004ya}. Following computations are performed by utilizing the
matrix elements
\begin{eqnarray}
&&\langle 0|J_{\nu }^{B_{c}^{\ast }}|B_{c}^{\ast +}(p^{\prime },\varepsilon
)\rangle =f_{B_{c}^{\ast }}m_{B_{c}^{\ast }}\varepsilon _{\nu }(p^{\prime }),
\notag \\
&&\langle 0|J^{B_{c}}|B_{c}^{-}(q)\rangle =\frac{f_{B_{c}}m_{B_{c}}^{2}}{%
m_{b}+m_{c}}.
\end{eqnarray}%
Here, the decay constant and polarization vector of $B_{c}^{\ast +}$ are
denoted by $f_{B_{c}^{\ast }}$ and $\varepsilon _{\nu }(p^{\prime })$,
respectively. The decay constant of the meson $B_{c}^{-}$ is labeled by $%
f_{B_{c}}$. The decay constants $f_{B_{c}^{\ast }}$ and $f_{B_{c}}$ are
chosen as $471~\mathrm{MeV}$ and $(371\pm 37)~\mathrm{MeV}$ \cite%
{Eichten:2019gig,Wang:2024fwc}, respectively. The vertex $\mathcal{M}_{%
\mathrm{AV}}B_{c}^{\ast +}B_{c}^{-}$ has the following form
\begin{eqnarray}
&&\langle B_{c}^{\ast +}(p^{\prime },\varepsilon )B_{c}^{-}(q)|\mathcal{M}_{%
\mathrm{AV}}(p,\epsilon )\rangle =g_{3}(q^{2})\left[ p\cdot q\epsilon \cdot
\varepsilon ^{\ast }\right.  \notag \\
&&\left. -p\cdot \varepsilon ^{\ast }q\cdot \epsilon \right] .
\end{eqnarray}%
Then $\Pi _{3\mu \nu }(p,p^{\prime })$ expressed in terms of the particles'
parameters becomes equal to
\begin{eqnarray}
&&\Pi _{3\mu \nu }^{\mathrm{Phys}}(p,p^{\prime })=\frac{g_{3}(q^{2})\Lambda
f_{B_{c}^{\ast }}m_{B_{c}^{\ast }}f_{B_{c}}m_{B_{c}}^{2}}{%
(m_{b}+m_{c})\left( p^{2}-m^{2}\right) \left( p^{\prime 2}-m_{B_{c}^{\ast
}}^{2}\right) }  \notag \\
&&\times \frac{1}{(q^{2}-m_{B_{c}}^{2})}\left[ \frac{m^{2}-m_{B_{c}^{\ast
}}^{2}+q^{2}}{2}g_{\mu \nu }-p_{\mu }p_{\nu }+p_{\mu }^{\prime }p_{\nu
}\right.  \notag \\
&&\left. -\frac{m^{2}}{m_{B_{c}^{\ast }}^{2}}p_{\mu }^{\prime }p_{\nu
}^{\prime }+\frac{m^{2}+m_{B_{c}^{\ast }}^{2}-q^{2}}{2m_{B_{c}^{\ast }}^{2}}%
p_{\mu }p_{\nu }^{\prime }\right] +\cdots .
\end{eqnarray}%
The function $\Pi _{3\mu \nu }(p,p^{\prime })$ computed using the heavy
quark propagators reads%
\begin{eqnarray}
&&\Pi _{3\mu \nu }^{\mathrm{OPE}}(p,p^{\prime })=-\frac{1}{2}\int
d^{4}xd^{4}ye^{ip^{\prime }y}e^{-ipx}\left\{ \mathrm{Tr}\left[ \gamma
_{5}S_{b}^{jb}(-x)\right. \right.  \notag \\
&&\left. \times \gamma _{\nu }S_{c}^{bj}(x)\right] \mathrm{Tr}\left[ \gamma
_{\mu }S_{c}^{ia}(y-x)\gamma _{5}S_{b}^{ai}(x-y)\right]  \notag \\
&&+\mathrm{Tr}\left[ \gamma _{5}S_{b}^{jb}(-x)\gamma _{5}S_{c}^{bj}(x)\right]
\mathrm{Tr}\left[ \gamma _{\mu }S_{c}^{ia}(y-x)\right.  \notag \\
&&\left. \left. \times \gamma _{\nu }S_{b}^{ai}(x-y)\right] \right\} .
\label{eq:CF4}
\end{eqnarray}

To derive the sum rule for $g_{3}(q^{2})$ we utilize invariant amplitudes
corresponding to the components proportional to $g_{\mu \nu }$ in $\Pi
_{3\mu \nu }^{\mathrm{Phys}}(p,p^{\prime })$ and $\Pi _{3\mu \nu }^{\mathrm{%
OPE}}(p,p^{\prime })$. Numerical analysis is done over the range $Q^{2}=2-30~%
\mathrm{GeV}^{2}$ using the following parameters: For $M_{2}^{2}$ and$\
s_{0}^{\prime }$ in the $B_{c}^{\ast +}$ channel, we employ
\begin{equation}
M_{2}^{2}\in \lbrack 6.5,7.5]~\mathrm{GeV}^{2},\ s_{0}^{\prime }\in \lbrack
50,51]~\mathrm{GeV}^{2}.
\end{equation}%
The extrapolating function $\mathcal{G}_{3}(Q^{2})$ has the parameters: $%
\mathcal{G}_{3}^{0}=0.34~\mathrm{GeV}^{-1}$, $a_{3}^{1}=1.82$, and $%
a_{3}^{2}=-12.12$. For the strong coupling $g_{3}$, we get
\begin{equation}
g_{3}\equiv \mathcal{G}_{3}(-m_{B_{c}}^{2})=(1.1\pm 0.2)\times 10^{-1}\
\mathrm{GeV}^{-1}.
\end{equation}%
The partial width of the decay $\mathcal{M}_{\mathrm{AV}}\rightarrow
B_{c}^{\ast +}B_{c}^{-}$ is equal to

\begin{equation}
\Gamma \left[ \mathcal{M}_{\mathrm{AV}}\rightarrow B_{c}^{\ast +}B_{c}^{-}%
\right] =(18.9\pm 6.0)~\mathrm{MeV}.  \label{eq:DW3}
\end{equation}

\bigskip
%%%%%%%%%%%%%%%%%%%%%%%%%%%%%%%%%%%%%%%%%%%%%%%%%%%%%%%%%%%%%%%%%%5

\subsection{$\mathcal{M}_{\mathrm{AV}}\rightarrow B_{c}^{\ast -}B_{c}^{+}$}

%%%%%%%%%%%%%%%%%%%%%%%%%%%%%%%%%%%%%%%%%%%%%%%%%%%%%%%%%%%%%%%%%%%%%%%

The process $\mathcal{M}_{\mathrm{AV}}\rightarrow B_{c}^{\ast -}B_{c}^{+}$
is studied by the same way. Omitting details we provide below our results
for the parameters of this decay.

The correlation function $\Pi _{4\mu \nu }^{\mathrm{OPE}}(p,p^{\prime })$
differs from Eq.\ (\ref{eq:CF4}) by $S_{c}(z)\leftrightarrow S_{c}(z)$
replacements. The strong coupling $g_{4}$ is equal to%
\begin{equation}
g_{4}\equiv \mathcal{G}_{4}(-m_{B_{c}}^{2})=(1.0\pm 0.2)\times 10^{-1}\
\mathrm{GeV}^{-1},
\end{equation}%
and is obtained by means of the extrapolating function $\mathcal{G}%
_{4}(Q^{2})$ with parameters $\mathcal{G}_{4}^{0}=0.28~\mathrm{GeV}^{-1}$, $%
a_{4}^{1}=1.41$, and $a_{4}^{2}=-12.29$. The partial width of this decay
amounts to
\begin{equation}
\Gamma \left[ \mathcal{M}_{\mathrm{AV}}\rightarrow B_{c}^{\ast -}B_{c}^{+}%
\right] =(15.5\pm 5.3)~\mathrm{MeV}.  \label{eq:DW4}
\end{equation}

Having used information about partial widths of the dominant channels of the
hadronic molecule $\mathcal{M}_{\mathrm{AV}}$, it is not difficult to
estimate its full decay width

\begin{equation}
\Gamma \left[ \mathcal{M}_{\mathrm{AV}}\right] =(93\pm 14)~\mathrm{MeV},
\end{equation}%
which means that $\mathcal{M}_{\mathrm{AV}}$ is relatively broad resonance.

%%%%%%%%%%%%%%%%%%%%%%%%%%%%%%%%%%%%%%%%%%%%%%%%%%%%%%%%%%%%%%

\section{Concluding notes}

\label{sec:Conc}

%%%%%%%%%%%%%%%%%%%%%%%%%%%%%%%%%%%%%%%%%%%%%%%%%%%%%%%%%%%

We have investigated the axial-vector hadronic molecule $\mathcal{M}_{%
\mathrm{AV}}=(B_{c}^{\ast +}B_{c}^{-}+B_{c}^{+}B_{c}^{\ast -})/2$ in the
framework of QCD sum rule method. Our result for the mass $m=(12770\pm 60)~%
\mathrm{MeV}$ of $\mathcal{M}_{\mathrm{AV}}$ gives important information
about this molecule. In fact, it is seen that the molecular structure $%
\mathcal{M}_{\mathrm{AV}}$ is unstable against two-meson fall-apart
processes.

As it has been noted above, the hadronic molecules $B_{c}^{(\ast
)+}B_{c}^{(\ast )-}$ were also explored in Ref.\ \cite{Liu:2023gla}. In the
case of the axial-vector structure $B_{c}^{\ast +}B_{c}^{-}+B_{c}^{\ast
-}B_{c}^{+}$ with cutoff momentum $\Lambda =800~\mathrm{MeV}$ the authors
fixed the pole at $E_{\mathrm{p}}=(12604.8-i0)~\mathrm{MeV}$ that can be
qualified as hadronic molecule with the mass $m=12604.8$. This prediction
smaller than one obtained in the present work. But, this molecule is still
unstable against strong dissociations to mesons $J/\psi \eta _{b}$ and $%
\Upsilon \eta _{c}$.

We have calculated the partial widths of $\mathcal{M}_{\mathrm{AV}}$
dissociations into $J/\psi \eta _{b}$, $\Upsilon \eta _{c}$, $B_{c}^{\ast
+}B_{c}^{-}$ and $B_{c}^{+}B_{c}^{\ast -}$ mesons which are dominant decay
modes of this molecular state. To calculate the widths of two-meson decays
we have applied the methods of the three-point sum rule approach. This is
required for estimation of strong couplings at various $\mathcal{M}_{\mathrm{%
AV}}$-meson-meson vertices. The full decay width of $\mathcal{M}$ saturated
by these decay modes is equal to $\Gamma \left[ \mathcal{M}_{\mathrm{AV}}%
\right] =(93\pm 14)~\mathrm{MeV.}$

In our previous publications \cite{Agaev:2024wvp,Agaev:2024mng,Agaev:2024qbh}%
, we carried out detailed analysis of the tetraquarks $bc\overline{b}%
\overline{c}$ with quantum numbers $0^{+}$, $1^{+}$ and $2^{+}$. \ We have
treated them as diquark-antidiquark compounds. The axial-vector state $T$
was modeled as a tetraquark composed of the scalar diquark and axial-vector
antidiquark components \cite{Agaev:2024mng}. The mass and full width of this
tetraquark were found equal to $m=(12715\pm 90)~\mathrm{MeV}$ and $\Gamma %
\left[ T\right] =(140\pm 13)~\mathrm{MeV}$, respectively.

Having compared the masses of these two axial-vector states, one sees that
the hadronic molecule $\mathcal{M}_{\mathrm{AV}}$ is heavier than the
diquark-antidiquark state though differences between their masses are small
and by taking into account uncertainties of computations, one may confirm
that they have very close masses.

Detailed investigations of the diquark-antidiquark and hadronic molecular
states $bc\overline{b}\overline{c}$ with various quantum numbers and
calculation their parameters are necessary to find processes where they may
be discovered.

\end{document}